\documentclass[aps,prl,twocolumn,groupedaddress]{revtex4}

\usepackage{amssymb,amsmath}
\usepackage{graphicx}
\usepackage{color}
\def\beq{\begin{equation}}
\def\eeq{\end{equation}}
\begin{document}


\title{Quantum simulation of the Klein paradox with trapped ions}


\author{R.~Gerritsma$^{1,2}$}
\author{B.~P.~Lanyon$^{1,2}$}
\author{G.~Kirchmair$^{1,2}$}
\author{F.~Z\"ahringer$^{1,2}$}
\author{C.~Hempel$^{1,2}$}
\author{J.~Casanova$^{3}$}
\author{J.~J.~Garc{\'i}a-Ripoll$^5$}
\author{E.~Solano$^{3,4}$}
\author{R.~Blatt$^{1,2}$}
\author{C.~F.~Roos$^{1,2}$}

\affiliation{$^1$Institut f\"ur Experimentalphysik, Universit\"at Innsbruck, Technikerstr.~25, A-6020 Innsbruck, Austria\\
$^2$Institut f\"ur Quantenoptik und Quanteninformation,
\"Osterreichische Akademie der Wissenschaften, Otto-Hittmair-Platz
1, A-6020 Innsbruck, Austria\\
$^3$ Departamento de Qu\'imica F\'isica, Universidad del Pa\'is Vasco - Euskal Herriko
Unibertsitatea, Apdo. 644, 48080 Bilbao, Spain\\
$^4$ IKERBASQUE, Basque Foundation for Science, Alameda Urquijo 36, 48011 Bilbao, Spain\\
$^5$ Instituto de F\'{\i}sica Fundamental, CSIC, Serrano 113-bis, 28006 Madrid, Spain}


\date{\today}

\begin{abstract}
We report on quantum simulations of relativistic scattering dynamics using trapped ions. The simulated state of a scattering particle is encoded in both the electronic and vibrational state of an ion, representing the discrete and continuous components of relativistic wave functions. Multiple laser fields and an auxiliary ion simulate the dynamics generated by the Dirac equation in the presence of a scattering potential. Measurement and reconstruction of the particle wave packet enables a frame-by-frame visualization of the scattering processes. By precisely engineering a range of external potentials we are able to simulate text book relativistic scattering experiments and study Klein tunneling in an analogue quantum simulator. We describe extensions to solve problems that are beyond current classical computing capabilities.


\end{abstract}


\maketitle


Simulating quantum mechanics using conventional computers rapidly becomes intractable as the physical systems to which it is applied grow larger. A proposed solution is to use highly controlled laboratory quantum systems themselves to perform such simulations~\cite{Feynman:1982,Lloyd:1996,Buluta:2009}. Systems currently under investigation for this purpose include photons~\cite{Lanyon:2010, Ma:2010}, trapped atoms~\cite{Greiner:2002,Trotzky:2008} and superconductors~\cite{Pritchett:2010}. A particularly promising approach uses trapped ions~\cite{Leibfried:2002,Porras:2004,Johanning:2009} with which several quantum simulations have recently been performed. One line of work is the simulation of quantum models of interacting spins~\cite{Porras:2004,Friedenauer:2008,Kim:2010}. Here, the internal states of ions encode the spin states and spin-spin interactions are simulated by laser induced state-dependent forces. Recently, a quantum simulation of the dynamics of a free relativistic particle has been performed in our group~\cite{Gerritsma:2010}. In this case, in contrast to simulations of spin systems, both discrete and continuous variables have to be simulated. In this paper we perform quantum simulations of the scattering dynamics of relativistic quantum particles. We show how a wide range of external potentials, from which the simulated particle can scatter, is engineered using two ions coupled via laser fields. We measure wave packets of the scattering particles, visualizing Klein tunneling `frame-by-frame'. Finally we describe extensions to efficiently simulate processes that are beyond current classical computing capabilities.

In its original form~\cite{Klein:1929} the Klein paradox considers a relativistic electron described by the Dirac equation, with total energy $E$ and rest mass energy $mc^2$, hitting a step-shaped potential barrier of height $V$. For barrier heights smaller than the kinetic energy of the electron, $V<E-mc^2$, the particle is predicted to partially transmit. For a slightly larger barrier, $E+mc^2>V>E-mc^2$, the particle should completely reflect. These situations agree with the predictions of the (non-relativistic) Schr{\"o}dinger equation. However, for $V> E+mc^2$ the particle can propagate undamped in the potential barrier, by turning into its anti-particle. This effect is known as the Klein paradox. Klein's results have been extended to other types of potentials~\cite{Sauter:1931,Galic:1988,Giachetti:2008} and the physics of Klein tunneling emerges for electrons in graphene~\cite{Katsnelson:2006,Neto:2009,Young:2009}. In quantum field theory, the paradox is resolved by the notion of pair creation by the external potential~\cite{Hund:1941,Schwinger:1951}.

\begin{figure}
\begin{center}
\includegraphics[width=8cm]{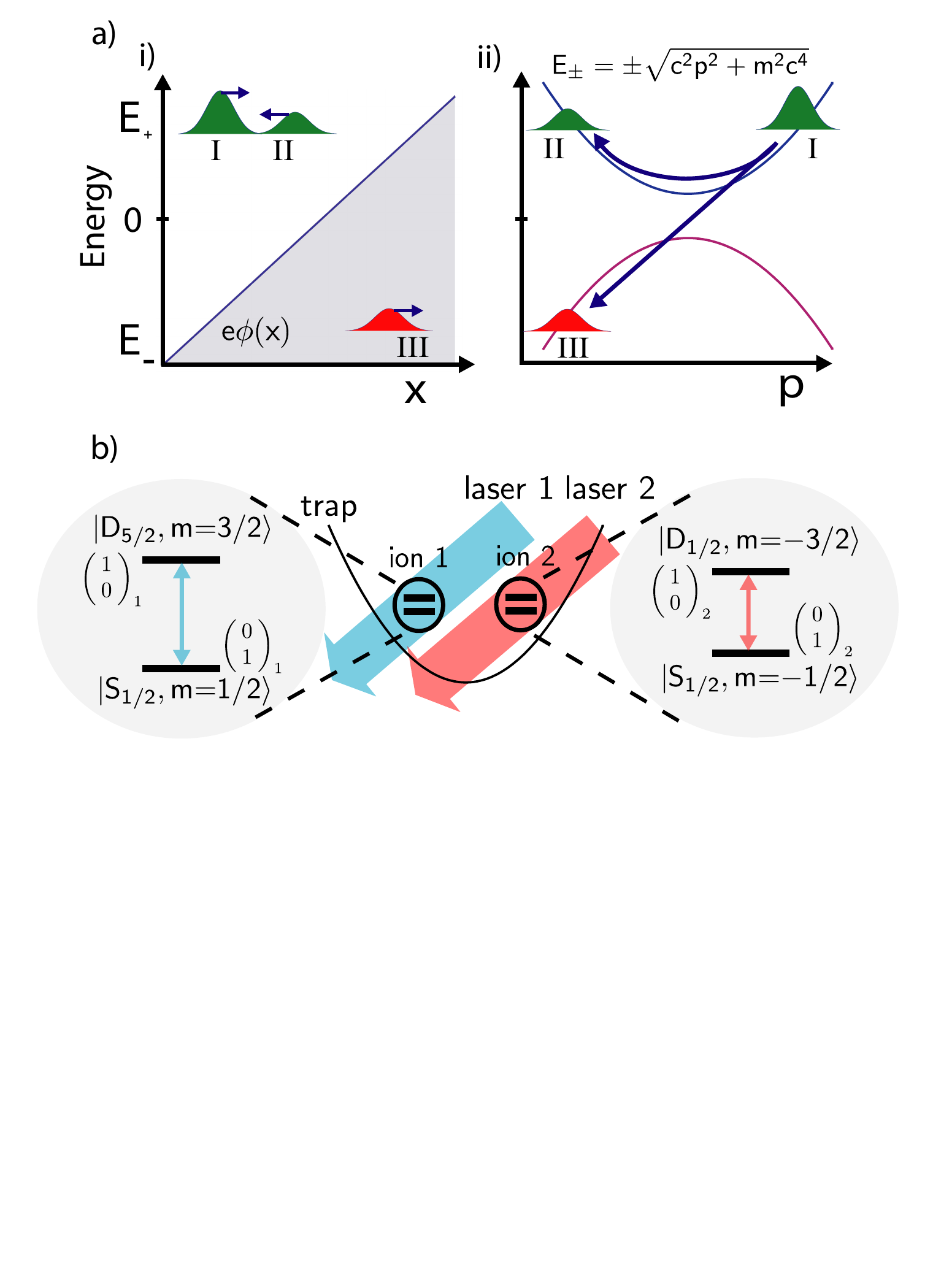}
\vspace{-50mm}
\caption{\label{fig1:setup}
(a) Klein tunneling in position i) and momentum space ii), for a relativistic particle scattering from a linear potential $\phi(x)=g x$. A wavepacket initially in the positive energy branch and with positive momentum (I) moves up a slope where it can reflect (II) while remaining in the positive energy branch or tunnel  (III) while switching energy branch. (b) The Dirac spinor is encoded in an internal state of ion~1 and a collective motional state of both ions. The Hamiltonian for a free Dirac particle ($H_d$) is implemented by bichromatic laser 1 which couples ion 1 to the collective motional mode. A second bichromatic laser couples ion 2 to the same motional mode and in this way creates the potential $\phi(x)$ (color online).}
\vspace{-9mm}
\end{center}
\end{figure}

In one dimension, the Dirac equation for a particle in an electrostatic potential $\phi(x)$ is given by\cite{Casanova:2010}

\beq
i\hbar\frac{\partial \Psi}{\partial t}=\left(c \,\hat{p}\,\sigma_x +mc^2\sigma_z +e\phi(x)I_2\right)\Psi.
\eeq

Here $c$ is the speed of light, $\hat{p}$ the momentum operator, $m$ the particle mass and $e$ its charge. The matrices $\sigma_j$, $j=x,y,z$ are the Pauli matrices and $I_2$ is the identity matrix. In one dimension there is no spin (no magnetic fields) and therefore the wave functions $\Psi$ are 2-component spinors, reflecting that there are positive and negative energy solutions  $E_\pm=\pm\sqrt{c^2p^2+m^2c^4}$. A spinor allows for arbitrary superpositions of these components. Free particles ($\phi(x)=0$) in one of the two energy branches remain there indefinitely, but for $\phi(x)\neq 0$ the spinor can switch energy branch and Klein tunneling can occur. For linear potentials $e\phi(x)=gx$ the situation is conceptually equivalent to Landau Zener tunneling~\cite{Sauter:1931,Casanova:2010,Wittig:2005}, as shown in Fig~\ref{fig1:setup}(a). Depending on the size of the splitting of the two energy branches ($2mc^2$) and the acceleration ($g/m$), the particle can either adiabatically follow the positive energy branch and reflect, or make a non-adiabatic transition to the negative energy branch and tunnel. For ultra-relativistic particles, the probability for tunneling is given by $P_{tunnel}=e^{-2\pi\Gamma}$, with $\Gamma=m^2c^3/2\hbar g$.

\begin{figure*}
\includegraphics[width=12cm]{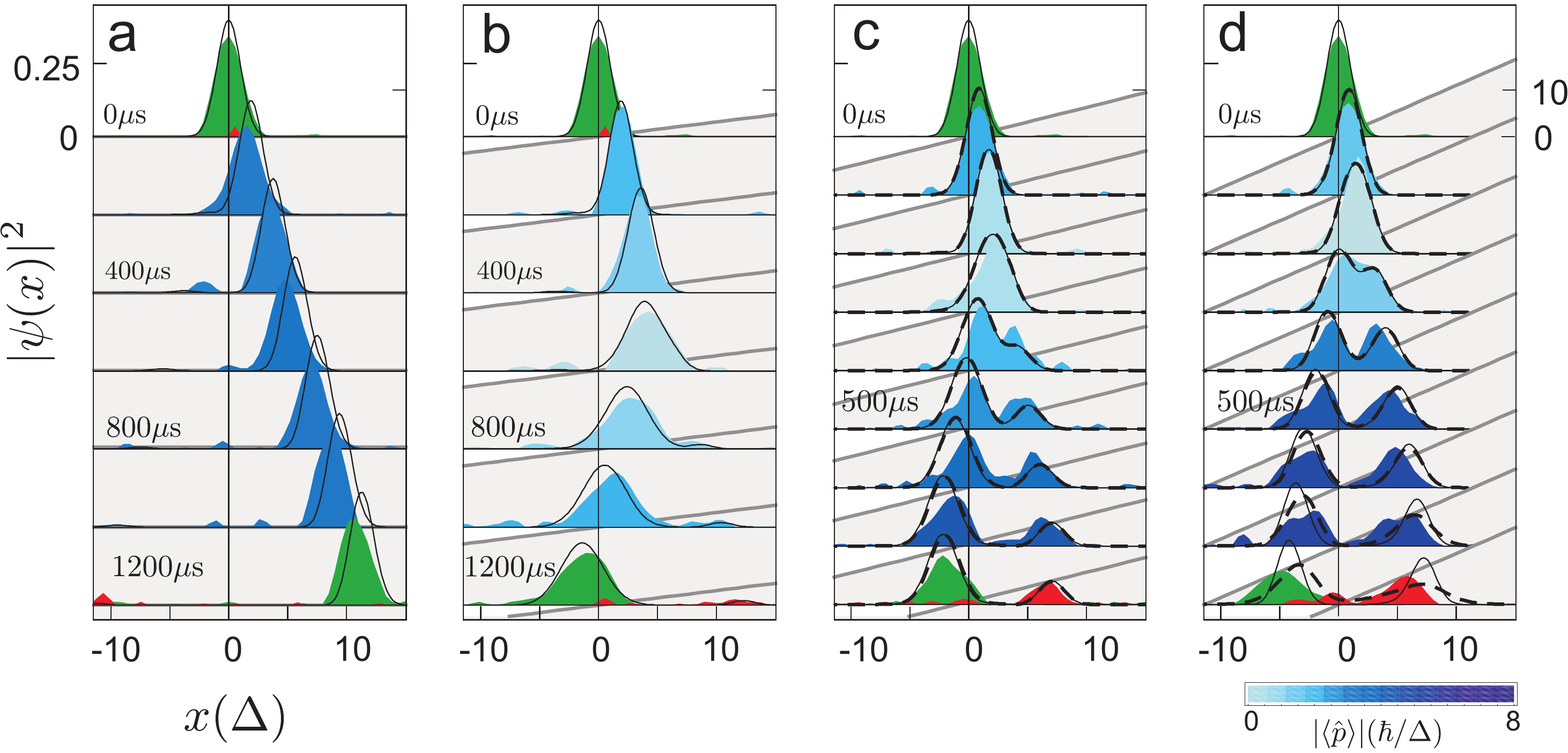}
\vspace{-17mm}
\caption{\label{fig2:slope}
Quantum simulation of relativistic scattering for linear potentials. Measured particle wave packets (filled curves) are compared with ideal predictions (solid black lines) and predictions taking corrections to the Lamb-Dicke approximation into account (dashed black lines).  In the first and last frames of each sequence the positive (green) and negative (red) energy components are reconstructed separately. The blue color scale of these panels represents the measured expectation value of momentum. The axis on the right shows the potential energy in units of initial kinetic energy. Without a potential, the particle moves to the right with constant velocity (a). For a shallow potential gradient, the particle is almost completely reflected (b) and for steeper gradients (c,d) part of the wave packet propagates into the repulsive potential via Klein tunneling (color online). }
\vspace{-5mm}
\end{figure*}

In our experiment, two $^{40}$Ca$^+$  ions are trapped in a linear Paul trap with trapping frequencies $\omega_{ax}=2\pi$~1.36~MHz axially, and $\omega_{rad}=2\pi$~3~MHz radially. A spinor is encoded by mapping the continuous position and momentum components to the quadratures of a vibrational mode, and the discrete components to two long-lived internal electronic states of ion 1 (Fig.~\ref{fig1:setup}(b)). Evolution under the free Dirac Hamiltonian is realised using a bichromatic light field~\cite{Lamata:2007,Gerritsma:2010} with an overall detuning coupling the electronic state of ion~1 to the vibrational mode via both red and blue sidebands. Within the Lamb-Dicke approximation, this interaction yields the Hamiltonian $H_d = 2 \eta \Delta \tilde{\Omega}_1 \sigma_x^{(1)} \hat{p}+\hbar \Omega_1 \sigma_z^{(1)}$, where $\eta=0.044$ is the Lamb-Dicke parameter, $\tilde{\Omega}_1$ is the bichromatic Rabi frequency and $2\Omega_1$ is the overall detuning. The momentum operator $\hat{p}=i\hbar(a^{\dag}{-}a)/2\Delta$, where $\Delta =\sqrt{\hbar/4\tilde{m}\omega_{ax}}\approx$~7~nm is the size of the ion's ground-state wave function, $\tilde{m}$ the ion's mass (not to be confused with the mass of the simulated particle) and $a^{\dag}$ ($a$) are the creation (annihilation) operator on the centre-of-mass vibrational mode. The notation $\sigma_j^{(n)}$ represents Pauli operators on the subspace of ion $n$. The mapping to the free Dirac Hamiltonian is complete by making the identifications $c:=2\eta\tilde{\Omega}_1\Delta$ and $mc^2:=\hbar\Omega_1$.

External potentials are simulated using a second ion (ion 2) and another bichromatic light field coupling ion 2 to the same vibrational mode but via a different electronic transition (Fig.~\ref{fig1:setup}(b)). With appropriately set phases the interaction Hamiltonian is given by $e\phi_{1} =g\sigma_x^{(2)}\hat{x}$ with $g=\hbar\eta\tilde{\Omega}_2/\Delta$~\cite{Casanova:2010}. Here, $\tilde{\Omega}_2$ is the Rabi frequency of the second bichromatic light field and $\hat{x}=(a^{\dag}+a)\Delta$ is the position operator. When ion 2 is prepared in an eigenstate of $\sigma_x^{(2)}$ this operator can be replaced by its +1 eigenvalue, reducing the interaction to a linear potential and $\Gamma=\Omega_1^2/(4\eta^2\tilde{\Omega}_1\tilde{\Omega}_2)$.

To reconstruct the spatial probability distribution of a spinor after the simulation, another bichromatic light pulse maps information about the position/momentum of the spinor onto the internal state of ion 2~\cite{Wallentowitz:1995,Gerritsma:2010}. A measurement of this state, as a function of the pulse length, can be used to reconstruct the probability distribution~(see reference~\cite{Zaehringer:2010} and its online material). In short, (1) after the simulation ion 2 is prepared in one of the states $\frac{1}{\sqrt{2}}\binom{1}{1}_2$ or $\binom{0}{1}_2$, (2) a bichromatic light field is used for a time $t_{probe}$ to implement a displacement operation of the form $U_{probe}=\exp(-ik\hat{x}\sigma_y^{(2)}/2)$, with $k\propto t_{probe}$. This operation causes the motional state to split in phase space along the momentum-axis. The resulting interference fringes in the observable $\sigma_z^{(2)}$, which can be measured by fluorescence detection, as a function of $t_{probe}$ then represent the fourier components of the original motional state $\sin (kx)$ or $\cos (kx)$, depending on which state ion 2 was prepared in. A fourier transform (3) of this data gives access to the spatial probability distribution. Before this measurement, the internal states of the two ions can be traced out by a series of laser pulses pumping the internal states into $\binom{0}{1}_1$ and $\binom{0}{1}_2$. For spinors with high momentum, it is also possible to obtain the wavepackets associated with positive and negative energy separately. In these cases the internal spinor states are not entangled with the motional state, and are given by the states $\binom{1}{\pm1}_1$. A $\pi/2$ pulse can be used to map either of these energy states to $\binom{1}{0}_1$ after which a short (200~$\mu$s) projective fluorescence measurement is done. Since the state $\binom{1}{0}_1$ scatters no photons during fluorescence detection, the motional state of the ions remains intact and can be analyzed afterwards. The cases where ion 1 is found in $\binom{0}{1}_1$ are discarded.

At the start of each experiment, laser cooling, optical pumping and coherent laser pulses, in a magnetic field of 6~G, prepare the ions in the axial center-of-mass mode ground state and internal states $\frac{1}{\sqrt{2}}\binom{1}{1}_1$ and $\frac{1}{\sqrt{2}}\binom{1}{1}_2$ (see Fig.~1.). We create an initial spinor state corresponding to a particle with $\langle \hat{x}\rangle=0$ and $p_0=\langle \hat{p}\rangle=3.5\hbar/\Delta$ by a displacement along the momentum quadrature. The generated spinor, $\psi(x;t=0)\propto e^{ip_0x/\hbar}e^{-\frac{x^2}{4\Delta^2}}\,\binom{1}{1}_1$, is comprised largely of positive energy: $|\langle\psi(x;t=0)|P^+|\psi(x;t=0)\rangle|^2>0.98$, where $P^{+}$ is the projector onto the positive energy state. The rest mass energy is set to $\hbar\Omega_1=\hbar\,2\pi$~1.3~kHz and $\tilde{\Omega}_1=2\pi$~17.5~kHz corresponding to an equivalent speed of light of $c\sim0.01\Delta/\mu$s. Fig.~\ref{fig2:slope} shows the scattering for different slope gradients ($g$), achieved by setting  $\tilde{\Omega}_2/2\pi=0,22,50$ and 76~kHz, corresponding to $P_{tunnel}=0, 0.03, 0.21$ and $0.36$ respectively. The measured tunnel probabilities are given by 0.017(7)~$\{0\}$, 0.10(1)~$\{0.07\}$, 0.32(2)~$\{0.22\}$ and 0.45(3)~\{0.39\}, here we put results obtained by numerical calculations in curly brackets. Animated versions of each of the cases can be found in the online material linked to the paper (EPAPS Document No. [{\it number will be inserted by publisher}]).

We are also able to create approximately quadratic electric potentials which, due to Klein tunneling, are non-confining in the relativistic limit~\cite{Giachetti:2008}. The quadratic potential is implemented experimentally by detuning the bichromatic beam on ion 2 by $2\Omega_2$, so that the coupling between this ion and the vibrational mode becomes $\hbar\eta\tilde{\Omega}_2\sigma_x^{(2)}\hat{x}/\Delta+\hbar\Omega_2\sigma_z^{(2)}$. In the limit of a large detuning, $\Omega_2 \gg \eta\tilde{\Omega}_2$ the effective interaction Hamiltonian becomes $e\phi_{2}=q\sigma_z^{(2)}\hat{x}^2$ with $q{=}\hbar\eta^2\tilde{\Omega}_2^2/(2\Delta^2\Omega_2)$~\cite{Casanova:2010}. Preparing ion 2 in the +1 eigenstate of $\sigma_{z}$ this reduces to a quadratic electric potential.

\begin{figure}
\begin{center}
\includegraphics[width=9cm]{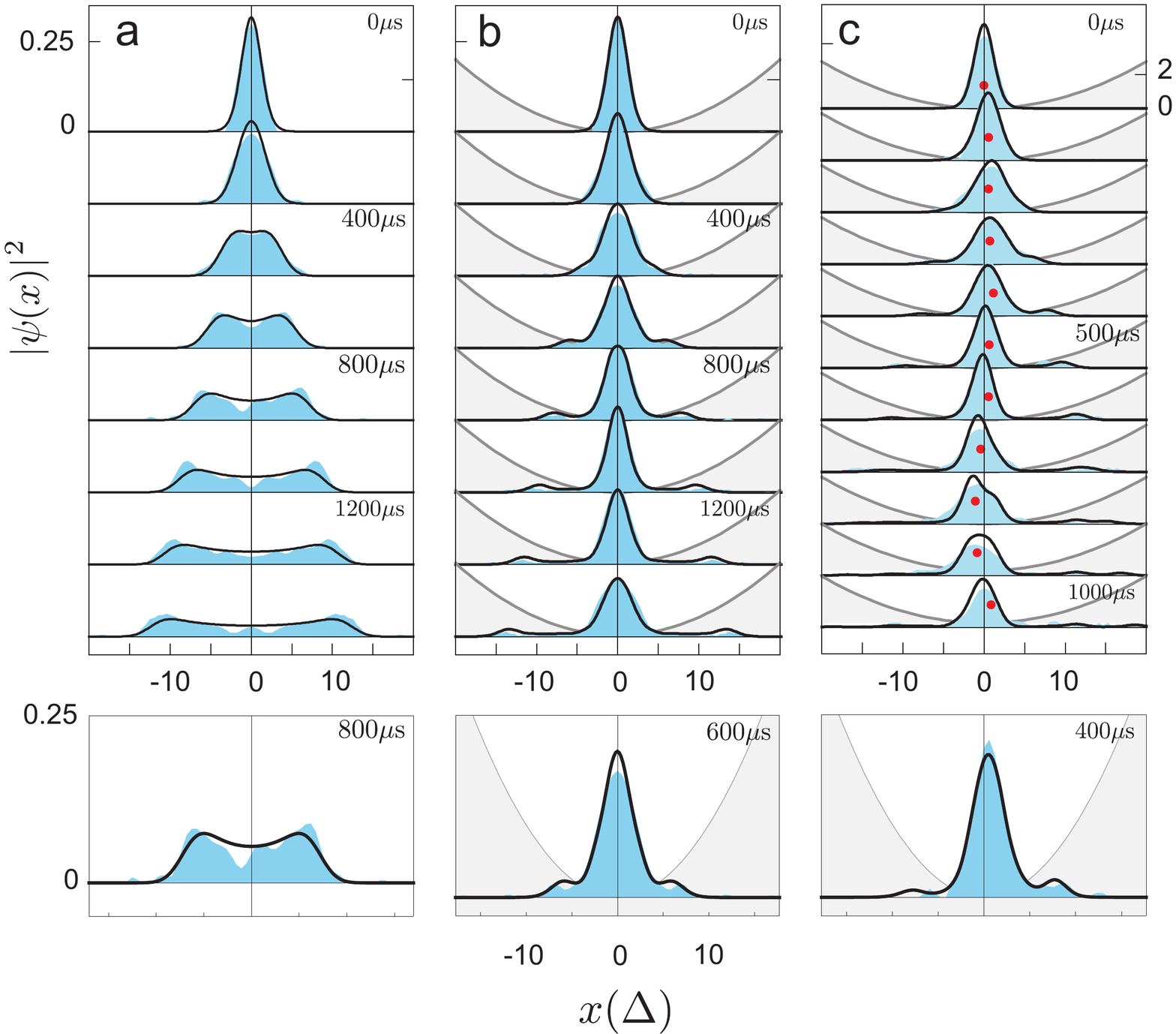}\caption{\label{fig3:quadratic}
Quantum simulation of scattering for quadratic potentials. Measured particle wave packets (filled curves) are compared with theoretical predictions (black lines) for quadratic potentials. The particle's rest mass energy is set to $\hbar\Omega_1=\hbar\,2\pi$~0.65~kHz. In (a) \& (b) the initial state has almost purely positive energy, and momentum $\langle \hat{p}\rangle=0\hbar/\Delta$. Without a potential (a) the wave packet diffuses. With the potential switched on (b) the spreading is significantly restricted, but can continue through Klein tunneling. In (c) the initial state has a small momentum $\langle \hat{p}\rangle=0.23\hbar/\Delta$ and the wavepacket oscillates. The red dots represent $\langle \hat{x}\rangle$. The lower three figures show larger versions of some of the wave packets. The axis on the right is in the same units as in Fig.~2. (color online)}
\end{center}
\vspace{-9mm}
\end{figure}

Figure~\ref{fig3:quadratic} shows results from our investigation of quadratic potentials. The first sequence (a) shows a particle, with initially positive energy, $\langle \hat{x} \rangle=0$ and $\langle \hat{p} \rangle=0$, evolving as a free particle ($\tilde{\Omega}_2=0$). As expected, the wave packet simply disperses.  Sequence (b) shows the same initial state evolving under a quadratic potential generated by setting $\Omega_2=2\pi$~33~kHz and $\tilde{\Omega}_2=2\pi$~50~kHz, such that $\eta\tilde{\Omega}_2/\Omega_2=0.067\ll 1$ and $q=2\pi$~73~Hz$~\hbar/\Delta^2$. The dynamics are clearly different in this case: the wavepacket is still unconfined, due to Klein tunneling, but it spreads more slowly. Sequence (c) shows results for the same potential, but for a positive energy particle with an initial momentum $\langle \hat{p} \rangle=0.23~\hbar/\Delta$. The scattering dynamics in this case correspond to that of a quantum relativistic mass on a spring that is given a small initial kick and show Klein tunneling at each turning point.

The state used for the simulations in Figs.~\ref{fig3:quadratic}(a) \& (b) was created by applying a bichromatic displacement operation to ion 1 of the form $U_{prep_2}=e^{-iH_{prep_2}t/\hbar}$  with $H_{prep_2}=\hbar\eta\Omega_{prep_2}\hat{x}\sigma_{x}^{(1)}/\Delta$ and ion 1 prepared in the internal state $\binom{1}{0}_2$. A 16~$\mu$s pulse with $\Omega_{prep_2}=2\pi$~83~kHz produces a spinor with $\langle \hat{p} \rangle=0$ and $|\langle\psi(x;t=0)|P^+|\psi(x;t=0)\rangle|^2>0.98$. The state used in Fig.~\ref{fig3:quadratic}~(c) was created in a similar way, but an additional pulse was used to give the state a momentum of $\langle \hat{p} \rangle =$~0.23~$\hbar/\Delta$.

There are a number of errors that reduce the quality of our quantum simulations. The internal state coherence time for the ions is $\approx$~3~ms, limited by magnetic field fluctuations, while the whole simulation takes up to 1.5~ms. The motional coherence is limited by slow drifts, which can change the trap frequency by about $\approx$~25~Hz. Both effects cause broadening of the wave packets, whereas systematic errors in state preparation due to slowly varying experimental settings can cause additional structure in the reconstructed data. For the steepest slope, states of more than 150 phonons on average were created, for which the Lamb-Dicke approximation starts to fail. This changes the simulated Hamiltonian somewhat, while also affecting the reconstruction~\cite{Zaehringer:2010}. All these errors can be improved by technological development.

Scaling up our quantum  simulation is conceptually straightforward: additional Dirac particles can be simulated by adding more ions and motional modes. Each extra spatial dimension requires an additional motional mode. Performing classical simulations at the same resolution achieved by the quantum simulations requires describing a 2~$\times$~200 dimensional Hilbert space (one qubit and 200 harmonic oscillator states). Adding just one Dirac particle increases the Hilbert space to $\approx16{\times}10^4$ dimensions, which would require 10$^3$~GB of memory just to store the Hamiltonian operator with 64-bit precision. Solving the scattering dynamics of two interacting Dirac particles is in general an open problem into which an ion trap quantum simulator could provide new insight. It is also possible to extend this work to simulate interactions with magnetic, scalar and much studied confining potentials such as the Dirac oscillator~\cite{Bermudez:2007}. Other possible extensions would be to simulate quantized Dirac fields or Majorana physics~\cite{Casanova:2010b}.

\begin{acknowledgments}
We gratefully acknowledge support by the Austrian Science Fund (FWF), by the European Commission (Marie-Curie program), by the Institut f\"ur Quanteninformation GmbH, IARPA. E.S. thanks support from UPV/EHU Grant GIU07/40, Spanish MICINN project FIS2009-12773-C02-01, Basque Government Grant IT472-10, EuroSQIP and SOLID European projects. J.C. acknowledges the Basque Government BFI08.211. J.J.G.-R. acknowledges the Spanish projects MICINN FIS2009-10061 and QUITEMAD.
\end{acknowledgments}

\end{document}